\documentclass[amsmath,11pt]{article}

\usepackage{amsfonts,amsmath,amssymb,graphics,epsfig}
\usepackage{inputenc,xspace,url}

\date{\empty}

\allowdisplaybreaks

\begin{document}

\title{\bf The Raychaudhuri equation in spacetimes with torsion and non-metricity}

\author{Damianos Iosifidis${}^1$, Christos G. Tsagas${}^{2,3}$ and Anastasios C. Petkou${}^1$\\ {\small ${}^1$\;Institute of Theoretical Physics}\\ {\small Department of Physics, Aristotle University of Thessaloniki,  Thessaloniki 54124, Greece}\\ {\small ${}^2$\;Section of Astrophysics, Astronomy and Mechanics}\\ {\small Department of Physics, Aristotle University of Thessaloniki, Thessaloniki 54124, Greece}\\ {\small $^3$DAMTP, Centre for Mathematical Sciences, University of Cambridge}\\ {\small Wilberforce Road, Cambridge CB3 0WA, UK}}

\maketitle

\begin{abstract}
We introduce and develop the 1+3 covariant approach to relativity and cosmology to spacetimes of arbitrary dimensions that have nonzero torsion and do not satisfy the metricity condition. Focusing on timelike observers, we identify and discuss the main differences between their kinematics and those of their counterparts living in standard Riemannian spacetimes. At the centre of our analysis lies the Raychaudhuri equation, which is the fundamental formula monitoring the convergence/divergence, namely the collapse/expansion, of timelike worldline congruences. To the best of our knowledge, we provide the most general expression so far of the Raychaudhuri equation, with applications to an extensive range of non-standard astrophysical and cosmological studies. Assuming that metricity holds, but allowing for nonzero torsion, we recover the results of analogous previous treatments. Focusing on non-metricity alone, we identify a host of effects that depend on the nature of the timelike congruence and on the type of the adopted non-metricity. We also demonstrate that in spaces of high symmetry one can recover the pure-torsion results from their pure non-metricity analogues, and vice-versa, via a simple ansatz between torsion and non-metricity.
\end{abstract}

\section{Introduction}\label{sI}
Given a manifold of arbitrary dimensions, one can measure distances between points and angles between vectors, once a metric has been introduced. On the other hand, for the parallel transport of vector and tensor fields on a manifold a connection is needed. In general, these two spacetime features, namely the metric and the connection, do not need to be related and (for the time being at least) there is no fundamental reason for them to do so, apart from simplicity. In classical general relativity, however, the metric and the (Levi-Civita) connection are related to each other, with the latter been expressed in terms of the former and its derivatives. More specifically, one arrives at the aforementioned relation after assuming that the metric is covariantly constant (aka metricity condition) and that the connection is symmetric (aka torsionless condition). Even though these two assumptions greatly simplify any theoretical analysis, we are not as yet aware of any fundamental mathematical, or physical, reason for selecting the Levi-Civita connection. The effort to identify alternative connections dates back to the work Weyl and Cartan towards the beginning of the last century~\cite{W} -- see also~\cite{S}. More specifically, Weyl considered torsionless spaces with non-metricity in an attempt to unify gravity with electromagnetism, whereas Cartan considered spaces with torsion. In the literature, the study of non-Riemannian contributions to gravity, is typically referred to as ``metric-affine gravity''~\cite{Sc}.

Motivated by the above, we extend the 1+3 covariant approach to general relativity and cosmology (see~\cite{TCM} for recent extensive reviews) to $n$-dimensional spacetimes that have nonzero torsion and do not satisfy the metricity condition. Our aim is to ``exploit'' the mathematical compactness and the geometrical/physical transparency of the covariant formalism in the ongoing quest for a deeper insight into these most general spacetimes. Torsion and non-metricity introduce new features to their host spaces. Among others, nonzero torsion implies that the Ricci curvature tensor and the matter energy-momentum tensor are no longer necessarily symmetric. This asymmetry could be seen as a generic spacetime feature, but it may also reflect the nonzero spin of its material content. Non-metricity, on the other hand, means that vectors and tensors do not maintain the same magnitude, as they are (parallelly) transported from one spacetime event to the next. As a result, the concepts of ``proper-length'' and ``proper time'' loose their conventional meaning when the metricity condition is violated. In view of these complications, in the first three chapters of this work we identify the key differences between our analysis and the standard treatments and also lay the foundations for extending the 1+3 formalism to general spacetimes with arbitrary dimensions, nonzero torsion and non-metricity.

At the centre of our study lies the Raychaudhuri equation, which has long been used to describe the mean kinematics of self-gravitating media (e.g.~see~\cite{Wa}). In particular, Raychaudhuri's formula has been at the core of the gravitational collapse studies and the related singularity theorems. Also, alternative versions of the same equation are currently  used in cosmology in search of an answer to the question posed by the recent universal acceleration. Here, we provide the most general (to the best of our knowledge) version of the Raychaudhuri equation, with no prior assumptions on the nature of the underlying gravitational theory. This ensures that our formula can be readily applied to a wide range of standard and non-standard astrophysical and cosmological problems. Assuming that metricity holds, but allowing for nonzero torsion, we find perfect agreement with the earlier 1+3 study of~\cite{PTB}. On the other hand, switching the torsion off and turning the non-metricity on reveals a rather intriguing resemblance between some (at least) of the torsion and the non-metricity effects. Motivated by this observation, as well as by analogous reports in the literature, we consider separately the simple cases of irrotational and shear free autoparallel congruences residing in empty (i.e.~Ricci-flat) spacetimes. In the first instance we assume nonzero torsion with metricity, while in the second we have non-metricity without torsion. Solving the Raychaudhuri equation in either case, we arrive at formally identical solutions. In particular, the pure-torsion solution can be recovered from its pure non-metricity counterpart (and vice versa) after imposing a surprisingly simple ansatz between these two spacetime features. We interpret this as clear demonstration of the so-called \textit{duality} between torsion and non-metricity (e.g.~see also~\cite{BSS}), which in spaces of high symmetry seems able to make the two theories phenomenologically identical.

\section{Spaces with torsion and non-metricity}\label{sSN-M}
Torsion and non-metricity modify the familiar Riemannian relations between the metric tensor, the connection and the curvature of the space. Here, we will briefly outline the main differences referring the reader to related reviews (e.g.~see~\cite{Sc}) for further discussion and details.

\subsection{Torsion and non-metricity tensors}\label{ssTN-MTs}
In the presence of torsion the connection of the space is generally asymmetric (i.e.~$\Gamma^{\mu}{}_{\nu\lambda}\neq \Gamma^{\mu}{}_{(\nu\lambda)}$), with its antisymmetric component giving the Cartan torsion tensor
\begin{equation}
S_{\mu\nu}{}^{\lambda}= \Gamma^{\lambda}{}_{[\mu\nu]}\,,  \label{tt}
\end{equation}
so that $S_{\mu\nu}{}^{\lambda}=S_{[\mu\nu]}{}^{\lambda}$ by default.\footnote{Round brackets denote symmetrisation, while square ones indicate antisymmetrisation.} At the same time, the metric is not necessarily covariantly conserved and the failure of the connection to do so is measured by the non-metricity tensor
\begin{equation}
Q_{\lambda\mu\nu}= -\nabla_{\lambda}g_{\mu\nu}\,,  \label{nmt}
\end{equation}
ensuring that $Q_{\lambda\mu\nu}=Q_{\lambda(\mu\nu)}$.\footnote{The non-metricity of the space implies that raising and lowering the indices are no longer trivial operations when covariant differentiation is involved. For instance, starting from (\ref{nmt}), one can show show that $\nabla^{\lambda}g^{\mu\nu}=Q^{\lambda\mu\nu}$. Also note that $\nabla^{\mu}=g^{\mu\nu}\nabla_{\nu}$ will define the contravariant counterpart of the covariant derivative operator.} The geometrical effect of torsion is that the parallel transport of a pair of vectors, along each other's direction, does not lead to a closed parallelogram. Non-metricity, on the other hand implies that the lengths of vectors are not preserved when they are parallelly transported in space.

Starting from the tensors defined above, one can construct two pairs of associated vectors. In particular, the torsion tensor leads to
\begin{equation}
S_{\mu}= S_{\mu\nu\lambda}g^{\nu\lambda}= S_{\mu\nu}{}^{\nu} \hspace{10mm} {\rm and} \hspace{10mm} \tilde{S}_{\mu}= \varepsilon_{\mu\nu\lambda\sigma} S^{\nu\lambda\sigma}\,,  \label{tvecs}
\end{equation}
where $\varepsilon_{\mu\nu\lambda\sigma}$ is the associated alternating tensor (with $\varepsilon_{\mu\nu\lambda\sigma}= \varepsilon_{[\mu\nu\lambda\sigma]}$). The former of these is the familiar torsion vector, while here we will refer to $\tilde{S}_{\mu}$ as the torsion pseudo-vector. The latter vanishes in highly symmetric spacetimes, like those associated with the familiar Friedmann-Robertson-Walker (FRW) models, because it leads to parity violation. For the non-metricity tensor, on the other hand, the related vectors are
\begin{equation}
Q_{\mu}= Q_{\mu\nu\lambda}g^{\nu\lambda}= Q_{\mu\nu}{}^{\nu} \hspace{10mm} {\rm and} \hspace{10mm} \tilde{Q}_{\mu}= g^{\lambda\nu}Q_{\lambda\nu\mu}= Q^{\nu}{}_{\nu\mu}\,,  \label{nmvecs}
\end{equation}
with $Q_{\mu}$ representing the so-called Weyl vector. Here onwards, we will refer to $\tilde{Q}_{\mu}$ as the second non-metricity vector.

\subsection{Special types of torsion and
non-metricity}\label{ssSTTN-M}
Following the above, the simplest types of torsion and non-metricity are of vector form. Then, in a $n$-dimensional space, the associated torsion and non-metricity tensors read
\begin{equation}
S_{\mu\nu}{}^{\lambda}= \frac{2}{n-1}\,S_{[\mu}\delta_{\nu]}{}^{\lambda} \hspace{10mm} {\rm and} \hspace{10mm} Q_{\mu\nu\lambda}= {1\over n}\, Q_{\mu}g_{\nu\lambda}\,,  \label{vectnm}
\end{equation}
respectively. Therefore, the torsion field is determined by the torsion vector ($S_{\mu}$) and the non-metricity by the Weyl vector ($Q_{\mu}$), in which case we are dealing with the so-called Weyl non-metricity. An additional interesting type of non-metricity is one that allows for fixed-length vectors, is which case the non-metricity tensor satisfies the constraint
\begin{equation}
Q_{\mu\nu\lambda}= v_{\mu}g_{\nu\lambda}- g_{\mu(\nu}v_{\lambda)}\,,  \label{flnm}
\end{equation}
where $v_{\mu}$ is an arbitrary vector field. In what follows, we will first consider the implications of torsion and non-metricity for the mean kinematics (i.e.~for the volume expansion/contraction scalar -- see \S~\ref{ssVSSVTs} below) of the host spacetime, without imposing any restrictions on either of these two geometrical features. Then, we will apply our generalised equations to some of the specific forms of torsion and non-metricity given in this section.\footnote{An additional simple form of torsion has $S_{\mu\nu\lambda}= \varepsilon_{\mu\nu\lambda\beta}\tilde{S}^{\beta}/3!$, where $\varepsilon_{\mu\nu\lambda\beta}$ is the 4-dimensional Levi-Civita tensor. Unlike vectorial torsion, however, this last form of torsion vanishes identically in spatially homogeneous and isotropic (FRW-type) cosmologies. Note that the latter spacetimes can naturally accommodate both the Weyl and the fixed-length forms of non-metricity (see \S~\ref{sCCs} below).}

\subsection{Curvature}\label{ssC}
As in conventional Riemannian geometry, the curvature of a space with torsion and non-metricity reflects the fact that the covariant differentiation is not a commutative operation. This is manifested in the Ricci identity, which applied to the contravariant vector $u^{\mu}$ reads
\begin{equation}
2\nabla_{[\mu}\nabla_{\nu]}u^{\lambda}= R^{\lambda}{}_{\beta\mu\nu}u^{\beta}+ 2S_{\mu\nu}{}^{\beta}\nabla_{\beta}u^{\lambda}\,,  \label{Ricci1}
\end{equation}
where $R^{\mu}{}_{\nu\lambda\beta}$ is the curvature tensor of the space given by
\begin{equation}
R^{\mu}{}_{\nu\lambda\beta}= 2\partial_{[\lambda}\Gamma^{\mu}{}_{|\nu|\beta}+ 2\Gamma^{\mu}{}_{\alpha[\lambda}\Gamma^{\alpha}{}_{|\nu|\beta]}\,.  \label{Riemann1}
\end{equation}
The above has only one symmetry, namely $R_{\mu\nu\lambda\beta}= -R_{\mu\nu\beta\lambda}$, in contrast to its purely Riemannian counterpart (i.e.~to the Riemann curvature tensor itself).

The reduced symmetries of the curvature tensor ensure that there are three independent contractions, namely
\begin{equation}
\tilde{R}_{\mu\nu}= g^{\lambda\beta}R_{\lambda\beta\mu\nu}= R^{\lambda}{}_{\lambda\mu\nu}\,,  \hspace{10mm} \hat{R}_{\mu\nu}= g^{\lambda\beta}R_{\mu\lambda\beta\nu}= R_{\mu}{}^{\lambda}{}_{\lambda\nu}\,,  \label{cntrRiemann}
\end{equation}
and
\begin{equation}
R_{\mu\nu}= g^{\lambda\beta}R_{\lambda\mu\beta\nu}= R^{\lambda}{}_{\mu\lambda\nu}\,,  \label{Riccit}
\end{equation}
respectively. The latter provides the familiar Ricci curvature tensor, while the former is usually referred to as the ``homothetic'' curvature tensor. One additional contraction leads to the Ricci scalar
\begin{equation}
R= g^{\mu\nu}R_{\mu\nu}\,,  \label{Riccis}
\end{equation}
which is uniquely defined (since $g^{\mu\nu}\tilde{R}_{\mu\nu}=0$ and $g^{\mu\nu}\hat{R}_{\mu\nu}=-R$).

An important for our purposes relation is obtained by applying the Ricci identity to the metric tensor of the host space. Assuming that the latter is $n$-dimensional with torsion and non-metricity, in addition to curvature, we arrive at
\begin{equation}
2\nabla_{[\mu}\nabla_{\nu]}g_{\lambda\beta}= -2R_{(\lambda\beta)\mu\nu}+ 2S_{\mu\nu}{}^{\alpha}\nabla_{\alpha}g_{\lambda\beta}\,.  \label{Ricci2}
\end{equation}
Expanding this expression and then using definition (\ref{nmt}) leads to
\begin{equation}
R_{(\mu\nu)\lambda\beta}= \nabla_{[\lambda}Q_{\beta]\mu\nu} -S_{\lambda\beta}{}^{\alpha}Q_{\alpha\mu\nu}\,,  \label{Riemann}
\end{equation}
which relates the curvature tensor with the torsion and the non-metricity tensors of the space.

\section{Spacetime splitting}\label{sS-TS}
The 1+3 covariant approach to relativity and cosmology decomposes the 4-D spacetime into one temporal and three spatial dimensions, while it utilises the Bianchi and the Ricci identities rather than the metric~\cite{TCM}. Over the years, this formalism has been extended to higher dimensions, and to spacetimes with nonzero torsion, but (to the best of our knowledge) it has never been applied to spaces where the metricity condition no longer applies (i.e.~when $\nabla_cg_{ab}\neq0$, with $g_{ab}$ being the spacetime metric). In what follows, we will attempt to take the first step in that direction.

\subsection{The timelike observes}\label{ssT-LOs}
In a $n$-dimensional spacetime, suppose that $u^{\mu}$, with $u^{\mu}={\rm d}x^{\mu}/{\rm d}\lambda$, is the $n$-velocity vector tangent to a congruence of timelike curves. The latter also define the worldlines of a family of observers, known as the fundamental observers. In the absence of metricity, the magnitude of the $n$-velocity vector is no longer preserved and for this reason it cannot be normalized to $-1$ (or in any other way). We may therefore write
\begin{equation}
u_{\mu}u^{\mu}= g_{\mu\nu}u^{\mu}u^{\nu}= -\ell^{2}\equiv -\phi(x^{\alpha})\,,  \label{ua2}
\end{equation}
where $\phi(x^{\alpha})$ is generally a function of both space and time.\footnote{Greek indices takes values from $0$ to $n-1$ and Latin indices run from $1$ to $n-1$ throughout this article.} As we will demonstrate throughout the rest of this manuscript, the spacetime dependence seen in Eq.~(\ref{ua2}) marks the starting point of a series of technical and conceptual differences between metric and non-metric cosmologies. To begin with, the affine parameter $\lambda$ does not necessarily coincide with the proper time ($\tau$) measured along the observers' timelike curves. In particular, setting ${\rm d}\tau^2=-g_{\mu\nu}{\rm d}x^{\mu}{\rm d}x^{\nu}$, applying the chain-rule of differentiation and employing (\ref{ua2}) we arrive at
\begin{equation}
{{\rm d}\tau\over{\rm d}\lambda}= \pm\ell\,,  \label{tau-lambda}
\end{equation}
with $\ell=\ell(x^{\alpha})$ due to the non-metricity of the spacetime. The above integrates to give the (nontrivial) relation $\tau=\pm\int l(x^{\alpha}){\rm d}\lambda+\mathcal{C}$ between the proper time measured along a timelike worldline and any affine parameter of that curve. Therefore, here onwards, we will use overdots to indicate differentiation with respect to the affine parameter (i.e.\;~$\dot{{}}={\rm d}/{\rm d}\lambda$) and primes to denote derivatives in terms of proper time (i.e.\;~${}^{\prime}={\rm d}/{\rm d}\tau$).\footnote{The coordinate time ($x^0$) measured by a comoving observers (those with $u^a=0$) relates to the affine parameter of their timelike worldlines by means of ${\rm d}x^0/{\rm d}\lambda=\pm\ell/\sqrt{-g_{00}}$. Then, setting $g_{00}=-1$ and using Eq.~(\ref{tau-lambda}) we deduce that ${\rm d}x^0={\rm d}\tau$. In other words, proper and coordinate time still coincide for comoving observers.}

\subsection{Temporal and spatial derivatives}\label{ssTSDs}
The non-metricity of the host spacetime also affects the (spatial) hypersurfaces orthogonal to the timelike $u^{\mu}$-field (e.g.~see~\cite{TCM} for a comparison). More specifically, the associated projection tensor is now given by
\begin{equation}
h_{\mu\nu}= g_{\mu\nu}+ {1\over\ell^2}\,u_{\mu}u_{\nu}\,,  \label{hab1}
\end{equation}
recalling that $g_{\mu\nu}g^{\mu\nu}=\delta_{\mu}{}^{\mu}=n$.
The above guarantees that $h_{\mu\nu}=h_{\nu\mu}$, that $h_{\mu\nu}u^{\mu}=0$ and that $h_{\mu\nu}h^{\mu\nu}=n-1$. In addition, following definition (\ref{hab1}), we obtain
\begin{equation}
h_{\mu\lambda}h^{\lambda\nu}= h_{\mu}{}^{\nu}= \delta_{\mu}{}^{\nu}+ {1\over\ell^2}\,u_{\mu}u^{\nu}\,.  \label{hab2}
\end{equation}

Overall, the timelike $n$-velocity field and the projector defined above, introduce an $1+(n-1)$ splitting of the spacetime into one temporal direction and $n-1$ spatial counterparts. We may therefore define the temporal and spatial derivatives of a general tensor field $T_{\alpha_1\cdots\alpha_n}^{\beta_1\cdots\beta_m}$ as
\begin{equation}
\dot{T}_{\alpha_1\cdots\alpha_n}^{\beta_1\cdots\beta_m}= u^{\mu}\nabla_{\mu}T_{\alpha_1\cdots\alpha_n}^{\beta_1\cdots\beta_m}  \label{tempD}
\end{equation}
and
\begin{equation}
{\rm D}_{\mu}T_{\alpha_1\cdots\alpha_n}^{\beta_1\cdots\beta_m}= h_{\mu}{}^{\lambda}h_{\alpha_1}{}^{\gamma_1}\cdots h_{\alpha_n}{}^{\gamma_n}h_{\delta_1}{}^{\beta_1}\cdots h_{\delta_m}{}^{\beta_m}\nabla_{\lambda}T_{\gamma_1\cdots \gamma_n}^{\delta_1\cdots\delta_m}\,,  \label{spD}
\end{equation}
respectively. On using the above, every spacetime variable, equation and operator can be decomposed into their temporal and spatial components.

\section{Kinematics}\label{sKs}
Torsion and non-metricity complicate considerably the kinematic description of the timelike observers introduced in the previous section. For example, some of the standard kinematic variables are no longer uniquely defined. Here, we will attempt to address these issues and also set up the mathematical formalism that we will use for the rest of our study.

\subsection{Path and hyper $n$-acceleration}\label{ssPHn-A}
The fact that the metric tensor is not covariantly conserved (i.e.~non-metricity) means that the processes of covariant differentiation and of index raising/lowering are not commutative. This in turn implies that there are two different $n$-acceleration vectors, namely a contravariant and a covariant one, defined by
\begin{equation}
A^{\mu}\equiv \dot{u}^{\mu}\equiv u^{\lambda}\nabla_{\lambda}u^{\mu}  \label{A}
\end{equation}
and
\begin{equation}
a_{\mu}\equiv \dot{u}_{\mu}\equiv  u^{\lambda}\nabla_{\lambda}u_{\mu}\,,  \label{a}
\end{equation}
respectively. Given that $\nabla^{\lambda}g
^{\mu\nu}= Q^{\lambda\mu\nu}$, with $Q^{\lambda\mu\nu}$ representing the non-metricity of the host spacetime, we deduce that $A^{\mu}\neq g^{\mu\nu}a_{\nu}$. More specifically, definitions (\ref{A}) and (\ref{a}) ensure that
\begin{equation}
A^{\mu}= a^{\mu}+ Q^{\nu\lambda\mu}u_{\nu}u_{\lambda}\,,  \label{Aa}
\end{equation}
in direct contrast to metric spacetimes where $A^{\mu}=a^{\mu}$~\cite{TCM}. It is then imperative to distinguish between these two types of $n$-acceleration. So, hereafter, we will name $A^{\mu}$ \textit{path $n$-acceleration}, since it vanishes along autoparallel trajectories/paths, while we will refer to $a_{\mu}$ as the \textit{hyper $n$-acceleration}, because it remains nonzero on autoparallel curves. In particular, Eq.~(\ref{Aa}) ensures that $a_{\mu}= -Q_{\lambda\mu\nu}u^{\lambda}u^{\nu}\neq0$ when $A_{\mu}=0$.\footnote{By definition, autoparallel curves have zero path acceleration, that is $A^{\mu}=\dot{u}^{\mu}= u^{\lambda}\nabla_{\lambda}u^{\mu}=0$. Autoparallel and geodesic trajectories coincide in Riemannian spaces, equipped with the Levi-Civita connection (i.e.~when $\Gamma^{\lambda}{}_{\mu\nu}= \Gamma^{\lambda}{}_{(\mu\nu)}$), but not in the presence of torsion (i.e.~when $\Gamma^{\lambda}{}_{[\mu\nu]}\neq0$), or non-metricity (i.e.~when $\nabla_{\lambda}g_{\mu\nu}\neq0$).}

An additional key difference between metric and non-metric spacetimes is that none of the two $n$-acceleration vectors defined above is normal to their associated $n$-velocity vector. Indeed, given that $u_{\mu}u^{\mu}=-\ell^2$, with $\ell=\ell(x^{\alpha})$, differentiating in terms of the affine parameter ($\lambda$ -- see \S~\ref{ssTSDs} before) leads to
\begin{equation}
A^{\mu}u_{\mu}= -{1\over2}\left(\ell^2\right)^{\cdot}+ {1\over2}\,Q_{\mu\nu\lambda}u^{\mu}u^{\nu}u^{\lambda}\,.  \label{Au}
\end{equation}
Similarly, recalling that $\nabla_{\lambda}g_{\mu\nu}= -Q_{\lambda\mu\nu}$, we arrive at
\begin{equation}
a^{\mu}u_{\mu}= -{1\over2}\left(\ell^2\right)^{\cdot}- {1\over2}\,Q_{\mu\nu\lambda}u^{\mu}u^{\nu}u^{\lambda}\,.  \label{au}
\end{equation}
The last two relations combine to give
\begin{equation}
\left(A^{\mu}+a^{\mu}\right)u_{\mu}= -\left(\ell^2\right)^{\cdot} \hspace{10mm} {\rm and} \hspace{10mm} \left(A^{\mu}-a^{\mu}\right)u_{\mu}= Q_{\mu\nu\lambda}u^{\mu}u^{\nu}u^{\lambda}\,.  \label{aux1}
\end{equation}
Finally, we should note that in the case autoparallel ``motion'' (i.e.~when $A_{\mu}=0$), expressions (\ref{aux1}a) and (\ref{aux1}b) guarantee that $\ell^2=\int Q_{\mu\nu\kappa}u^{\mu} u^{\nu}u^{\kappa}{\rm d}\lambda+\mathcal{C}$.

\subsection{Volume scalar, shear and vorticity 
tensors}\label{ssVSSVTs}
The irreducible kinematics of the $u_a$-field are determined by decomposing the associated covariant derivative into its temporal and spatial components, according to
\begin{equation}
\nabla_{\nu}u_{\mu}= {\rm D}_{\nu}u_{\mu}- {1\over\ell^2}\,(u_{\mu}\xi_{\nu}+a_{\mu}u_{\nu})- {1\over\ell^4}\,(u^{\lambda}a_{\lambda})u_{\mu}u_{\nu}\,,  \label{cNablau1}
\end{equation}
where ${\rm D}_{\nu}u_{\mu}=h_{\nu}{}^{\phi}h_{\mu}{}^{\lambda} \nabla_{\phi}u_{\lambda}$ (see definition (\ref{spD})). Also, $\xi_{\mu}=u^{\nu}\nabla_{\mu}u_{\nu}$ by definition with $\xi_{\mu}u^{\mu}=a_{\mu}u^{\mu}$ by construction.\footnote{By construction we have $\xi_{\mu}=u^{\nu}\nabla_{\mu}u_{\nu}$ and $\xi^{\mu}=u^{\nu}\nabla^{\mu}u_{\nu}$. Note, however, that the non-metricity of the space guarantees that $u^{\nu}\nabla_{\mu}u_{\nu}\neq u_{\nu}\nabla_{\mu}u^{\nu}$ in general, since $u^{\nu}\nabla_{\mu}u_{\nu}- u_{\nu}\nabla_{\mu}u^{\nu}=-Q_{\mu\nu\lambda}u^{\nu}u^{\lambda}$.} Moreover, the projected covariant derivative decomposes further into
\begin{equation}
{\rm D}_{\nu}u_{\mu}= {1\over n-1}\,\left(\Theta+{1\over\ell^2}\, a_{\lambda}u^{\lambda}\right)h_{\mu\nu}+ \sigma_{\mu\nu}+ \omega_{\mu\nu}\,,  \label{Du}
\end{equation}
with
\begin{equation}
\Theta= g^{\mu\nu}\nabla_{\nu}u_{\mu}= {\rm D}^{\mu}u_{\mu}- {1\over\ell^2}\, a_{\mu}u^{\mu}\,,  \label{Theta}
\end{equation}
representing a uniquely defined ``volume'' scalar (where ${\rm D}^{\mu}u_{\mu}=h^{\mu\nu}\nabla_{\nu}u_{\mu}$).\footnote{In the absence of metricity ${\rm D}^{\mu}u_{\mu}\neq{\rm D}_{\mu}u^{\mu}$, which implies that the ``spatial'' divergence of the $u_{\mu}$-field is not uniquely defined. More specifically, using definitions (\ref{nmt}), (\ref{nmvecs}b), (\ref{A}) and (\ref{a}), recalling that $Q_{\mu\nu\lambda}= Q_{\mu(\nu\lambda)}$ and employing the auxiliary relation (\ref{aux1}b), we find that
\begin{equation}
{\rm D}^{\mu}u_{\mu}= {\rm D}_{\mu}u^{\mu}- {1\over\ell^2}\,Q_{\mu\nu\lambda}u^{\mu}u^{\nu}u^{\lambda}- \tilde{Q}_{\mu}u^{\mu}\,.  \label{DuDu}
\end{equation}
The above explain/justify our choice of the uniquely defined $\Theta= g^{\mu\nu}\nabla_{\nu}u_{\mu}$ for the volume scalar. We should also point out that $\Theta\neq\nabla^{\mu}u_{\mu}$, since the divergence of the $n$-velocity is also not uniquely defined (i.e.~$\nabla^{\mu}u_{\mu}\neq\nabla_{\mu}u^{\mu}$)} When the latter is positive, the curves tangent to the $u_{\mu}$-field move apart and we have expansion. In the opposite case, on the other hand, the curves approach each other and there is contraction. Also, the variables
\begin{equation}
\sigma_{\mu\nu}= {\rm D}_{\langle\nu}u_{\mu\rangle} \hspace{10mm} {\rm and} \hspace{10mm} \omega_{\mu\nu}={\rm D}_{[\nu}u_{\mu]}\,,  \label{shrvrt}
\end{equation}
define the shear tensor and the vorticity tensor respectively.\footnote{Angled brackets indicate the symmetric and trace-free part of a second-rank tensor. For instance, the shear tensor is constructed as $\sigma_{\mu\nu}= h_{(\nu}{}^{\beta}h_{\mu)}{}^{\lambda} \nabla_{\beta}u_{\lambda}-[\Theta/(n-1)]h_{\mu\nu}$.} The former monitors kinematic anisotropies, namely ``shape'' distortions under constant ``volume'', while a nonzero vorticity implies that the $u_{\mu}$-field rotates.\footnote{Each of the three kinematic variables splits in its Riemannian and non-Riemannian parts. For instance the expansion scalar decomposes as $\Theta=\bar{\Theta}+ \Big(\tilde{Q}_{\mu}-Q_{\mu}/2-2S_{\mu}\Big)u^{\mu}$, with $\bar{\Theta}$ representing the Riemannian component.} Note that by construction $\sigma_{\mu}{}^{\mu}=0= \omega_{\mu}{}^{\mu}$ and $\sigma_{\mu\nu}u^{\nu}=0= \omega_{\mu\nu}u^{\nu}$. In other words, both the shear and the vorticity ``live'' in the observers ($n-1$)-dimensional rest-space. Finally, expressions (\ref{cNablau1}) and (\ref{Du}), combine to the following decomposition
\begin{equation}
\nabla_{\nu}u_{\mu}= {1\over n-1}\,\left(\Theta+{1\over\ell^2}\, a_{\lambda}u^{\lambda}\right)h_{\mu\nu}+ \sigma_{\mu\nu}+ \omega_{\mu\nu}- {1\over\ell^2}\,(u_{\mu}\xi_{\nu}+a_{\mu}u_{\nu})- {1\over\ell^4}\,(u^{\lambda}a_{\lambda})u_{\mu}u_{\nu}\,,  \label{cNablau2}
\end{equation}
of the covariant form ($\nabla_{\nu}u_{\mu}$) of the $n$-velocity gradient into the irreducible kinematic variables of the motion.

Given that $g^{\nu\beta}g^{\mu\lambda}\nabla_{\beta}u_{\lambda}= \nabla^{\nu}(g^{\mu\lambda}u_{\lambda})- u_{\lambda}\nabla^{\nu}g^{\mu\lambda}$ and recalling that $\nabla^{\lambda}g^{\mu\nu}=Q^{\lambda\mu\nu}$ -- see footnote~2 in S~\ref{ssTN-MTs}), one can show that the contravariant form ($\nabla^{\nu}u^{\mu}$) of the velocity gradient accepts the following irreducible decomposition
\begin{eqnarray}
\nabla^{\nu}u^{\mu}&=& {1\over n-1}\,\left(\Theta+{1\over\ell^2}\, a_{\lambda}u^{\lambda}\right)h^{\mu\nu}+ \sigma^{\mu\nu}+ \omega^{\mu\nu}- {1\over\ell^2}\,(u^{\mu}\xi^{\nu}+a^{\mu}u^{\nu})- {1\over\ell^4}\,(a_{\lambda}u^{\lambda}) u^{\mu}u^{\nu} \nonumber\\ &&+Q^{\nu\mu\lambda}u_{\lambda}\,, \label{ctNablau}
\end{eqnarray}
where $\Theta=g^{\mu\nu}\nabla_{\nu}u_{\mu}$ as in Eq.~(\ref{cNablau2}) above. Also, $\sigma^{\mu\nu}= g^{\mu\lambda}g^{\nu\beta}\sigma_{\lambda\beta}$ and $\omega^{\mu\nu}=g^{\mu\lambda}g^{\nu\beta}\omega_{\lambda\beta}$ are the contravariant components of the of the shear and the vorticity tensors respectively. Note, however, that $\sigma^{\mu\nu}\neq{\rm D}^{\langle\nu}u^{\mu\rangle}$ and $\omega^{\mu\nu}\neq{\rm D}^{[\nu}u^{\mu]}$ due to the non-metricity of the spacetime.

\section{The Raychaudhuri equation}\label{sRE}
The Raychaudhuri equation monitors the expansion, or the contraction, of a self-gravitating medium. It plays a fundamental role both in astrophysics and in cosmology and has been at the centre of all the singularity theorems. In what follows we will provide an expression for Raychaudhuri's formula in $n$-dimensional spaces with torsion and non-metricity.\footnote{Versions of the Raychaudhuri equation in spacetimes with nonzero torsion and/or spin have a fairly long history in the literature (e.g.~see~\cite{T} for a representative list). Here we adopt the formalism developed in~\cite{PTB}. Recently, there was also an attempt to extend Raychadhuri's formula to spaces with Weyl geometry~\cite{LBR}.}

\subsection{Deriving Raychaudhuri's formula}\label{ssDRF}
Raychaudhuri's formula is purely geometrical by nature and follows from a set of (also purely geometrical) relations, known as the Ricci identities. Applied to the $n$-velocity vector $u_{\mu}$ defined in \S~\ref{ssT-LOs}, the latter read
\begin{equation}
2\nabla_{[\mu}\nabla_{\nu]}u_{\lambda}= -R_{\beta\lambda\mu\nu}u^{\beta}+ 2S_{\mu\nu}{}^{\beta}\nabla_{\beta}u_{\lambda}\,,  \label{Ricci3}
\end{equation}
with $S_{\mu\nu\lambda}$ representing the torsion tensor and $R_{\mu\nu\lambda\beta}$ being the curvature tensor of the spacetime (so that $R_{\mu\nu\lambda\beta}=R_{\mu\nu[\lambda\beta]}$ -- see \S~\ref{ssTN-MTs} and \S~\ref{ssC} earlier). Contracting (\ref{Ricci3}) along $g^{\lambda\nu}u^{\mu}$ gives
\begin{equation}
g^{\lambda\nu}u^{\mu} \left(\nabla_{\mu}\nabla_{\nu}u_{\lambda} -\nabla_{\nu}\nabla_{\mu}u_{\lambda}\right)= -R_{\beta\lambda\mu\nu}u^{\beta}u^{\mu}g^{\lambda\nu}+ 2S_{\mu}{}^{\nu\lambda}u^{\mu}\nabla_{\lambda}u_{\nu}\,.  \label{Ray1}
\end{equation}
where the velocity gradient $\nabla_{\nu}u_{\mu}$ satisfies decomposition (\ref{cNablau2}). Using the latter, recalling that $\Theta=g^{\mu\nu}\nabla_{\nu}u_{\mu}$ and $\nabla_{\mu}Q^{\nu\lambda}=Q_{\mu}{}^{\nu\lambda}$ (see \S~\ref{ssTN-MTs} earlier), while employing definition (\ref{nmvecs}a) together with the symmetry property $Q_{\mu\nu\lambda}=Q_{\mu(\nu\lambda)}$ of the non-metricity tensor, the first term on the left-hand side of the above evaluates to
\begin{eqnarray}
g^{\lambda\nu}u^{\mu}\nabla_{\mu}\nabla_{\nu}u_{\lambda}&=& \dot{\Theta}- {1\over n-1} \left(\Theta+{1\over\ell^2}\,a_{\nu}u^{\nu}\right)Q_{\mu}u^{\mu}+ {1\over\ell^2}\,Q_{\mu\nu\lambda}u^{\mu}u^{\nu} (a^{\lambda}+\xi^{\lambda})- Q_{\mu\nu\lambda}u^{\mu}\sigma^{\nu\lambda} \nonumber\\ &&-{1\over\ell^2(n-1)} \left(\Theta-{{n-2}\over\ell^2}\,a_{\beta}u^{\beta}\right) Q_{\mu\nu\lambda}u^{\mu}u^{\nu}u^{\lambda}\,.  \label{lhsRay1}
\end{eqnarray}
Employing decompositions (\ref{cNablau2}) and (\ref{ctNablau}), while keeping in mind that $h_{\mu\nu}h^{\mu\nu}=n-1$, that $h_{\mu\nu}u^{\nu}=0=\sigma_{\mu\nu}u^{\nu}=\omega_{\mu\nu}u^{\nu}$, that $\sigma_{\mu\nu}h^{\mu\nu}=0=\omega_{\mu\nu}h^{\mu\nu}= \sigma_{\mu\nu}\omega^{\mu\nu}$, that $\xi_{\mu}u^{\mu}= a_{\mu}u^{\mu}$ and also using definition (\ref{nmvecs}b), the second term on the left-hand side of (\ref{Ray1}) becomes
\begin{eqnarray}
g^{\lambda\nu}u^{\mu}\nabla_{\nu}\nabla_{\mu}u_{\lambda}&=& -{1\over n-1}\,\Theta^2- 2\left(\sigma^2-\omega^2\right)+ {\rm D}^{\mu}a_{\mu}+ {1\over\ell^2}\,a_{\mu}A^{\mu}- {1\over\ell^2}\,\left(a_{\mu}u^{\mu}\right)^{\cdot}- {2\Theta\over\ell^2(n-1)}\,a_{\mu}u^{\mu} \nonumber\\ &&+{n-2\over\ell^4(n-1)}\left(a_{\mu}u^{\mu}\right)^2+ {2\over\ell^2}\,a_{\mu}\xi^{\mu}- {1\over n-1} \left(\Theta+{1\over\ell^2}\,a_{\beta}u^{\beta}\right) \tilde{Q}_{\mu}u^{\mu} \nonumber\\  &&-{1\over\ell^2(n-1)} \left(\Theta-{n-2\over\ell^2}\,a_{\beta}u^{\beta}\right) Q_{\mu\nu\lambda}u^{\mu}u^{\nu}u^{\lambda}- Q_{\mu\nu\lambda} \left(\sigma^{\mu\nu}+\omega^{\mu\nu}\right)u^{\lambda} \nonumber\\ &&+{1\over\ell^2}Q_{\mu\nu\lambda}\left(u^{\mu}\xi^{\nu} +a^{\mu}u^{\nu}\right)u^{\lambda}\,.  \label{lhsRay2}
\end{eqnarray}
Note that the scalars $\sigma^2=\sigma_{\mu\nu}\sigma^{\mu\nu}/2$ and $\omega^2=\omega_{\mu\nu}\omega^{\mu\nu}/2$ measure the magnitude of the shear and the vorticity tensors respectively.\footnote{In deriving expression (\ref{lhsRay2}) we have also used the auxiliary relation
\begin{equation}
\nabla^{\mu}a_{\mu}= {\rm D}^{\mu}a_{\mu}+ {1\over\ell^2}\,A^{\mu}a_{\mu}- {1\over\ell^2}\,\left(a_{\mu}u^{\mu}\right)^{\cdot}\,.  \label{auxRay}
\end{equation}}

Let us now turn our attention to the right-hand side of Eq.~(\ref{Ray1}). Starting from relation (\ref{Riemann}) that was obtained in \S~\ref{ssC} earlier, while recalling that $Q_{\mu\nu\lambda}= Q_{\mu(\nu\lambda)}$ and $S_{\mu\nu\lambda}=S_{[\mu\nu]\lambda}$, the first term on the right-hand side of expression (\ref{Ray1}) reads
\begin{eqnarray}
R_{\beta\lambda\mu\nu}u^{\beta}u^{\mu}g^{\lambda\nu}&=& R_{\mu\nu}u^{\mu}u^{\nu}+ \dot{\tilde{Q}}_{\mu}u^{\mu}- u^{\mu}u^{\nu}\nabla^{\lambda}Q_{\mu\nu\lambda}- Q_{\mu}{}^{\lambda\beta}Q_{\beta\lambda\nu}u^{\mu}u^{\nu} \nonumber\\ &&-2S_{\mu}{}^{\lambda\beta}Q_{\beta\lambda\nu}u^{\mu}u^{\nu}\,,  \label{rhsRay1}
\end{eqnarray}
with $R_{\mu\nu}=g^{\lambda\beta}R_{\lambda\mu\beta\nu}$ defining the Ricci curvature tensor. In addition, substituting decomposition (\ref{cNablau2}) and putting together definition (\ref{tvecs}a) and the symmetry property $S_{\mu\nu\lambda}=S_{[\mu\nu]\lambda}$ of the torsion tensor, the second term on the right-hand side of (\ref{Ray1}) recasts into
\begin{eqnarray}
S_{\mu}{}^{\nu\lambda}u^{\mu}\nabla_{\lambda}u_{\nu}&=& {1\over n-1} \left(\Theta+{1\over\ell^2}\,a_{\nu}u^{\nu}\right)S_{\mu}u^{\mu}+ S_{\mu\nu\lambda}u^{\mu}(\sigma^{\nu\lambda}+\omega^{\nu\lambda}) \nonumber\\ &&+{1\over\ell^2}\,S_{\mu\nu\lambda}a^{\mu}u^{\nu}u^{\lambda}\,.  \label{rhsRay2}
\end{eqnarray}

Finally, combining the intermediate relations (\ref{lhsRay1}), (\ref{lhsRay2}), (\ref{rhsRay1}) and (\ref{rhsRay2}), we obtain the generalisation of the Raychaudhuri equation to $n$-dimensional spaces with torsion and non-metricity, in addition to curvature, namely
\begin{eqnarray}
\dot{\Theta}&=& -{1\over n-1}\,\Theta^2- R_{\mu\nu}u^{\mu}u^{\nu}- 2\left(\sigma^2-\omega^2\right)+ {\rm D}^{\mu}a_{\mu}+ {1\over\ell^2}\,a_{\mu}A^{\mu} \nonumber\\ &&+{2\over n-1} \left(\Theta+{1\over\ell^2}\,a_{\nu}u^{\nu}\right)S_{\mu}u^{\mu}+ 2S_{\mu\nu\lambda}u^{\mu}(\sigma^{\nu\lambda}+\omega^{\nu\lambda})+ {2\over\ell^2}\,S_{\mu\nu\lambda}a^{\mu}u^{\nu}u^{\lambda} \nonumber\\ &&-{1\over\ell^2}(a_{\mu}u^{\mu})^{\cdot}- {2\Theta\over\ell^2(n-1)}\,a_{\mu}u^{\mu}+ {n-2\over\ell^4(n-1)}(a_{\mu}u^{\mu})^2+ {2\over\ell^2}\,a_{\mu}\xi^{\mu}- \dot{\tilde{Q}}_{\mu}u^{\mu} \nonumber\\ &&+{1\over n-1} \left(\Theta+{1\over\ell^2}\,a_{\nu}u^{\nu}\right) (Q_{\mu}-\tilde{Q}_{\mu})u^{\mu}-Q_{\mu\nu\lambda} (\sigma^{\mu\nu}+\omega^{\mu\nu}) u^{\lambda}- {1\over\ell^2}\,Q_{\mu\nu\lambda}u^{\mu}u^{\nu} (a^{\lambda}+\xi^{\lambda}) \nonumber\\ &&+Q_{\mu\nu\lambda}u^{\mu}\sigma^{\nu\lambda}+ {1\over\ell^2}\,Q_{\mu\nu\lambda}(u^{\mu}\xi^{\nu}+a^{\mu}u^{\nu}) u^{\lambda}+ u^{\mu}u^{\nu}\nabla^{\lambda}Q_{\mu\nu\lambda}+ Q_{\mu}{}^{\lambda\beta}Q_{\beta\lambda\nu}u^{\mu}u^{\nu} \nonumber\\  &&+2S_{\mu}{}^{\lambda\beta} Q_{\beta\lambda\nu}u^{\mu}u^{\nu}\,.  \label{nmtRay}
\end{eqnarray}
Note that only the terms in the first line on the right-hand side of the above have Riemannian analogues. More specifically, in the absence of torsion and in the presence of metricity (i.e.~when $S_{\mu\nu\lambda}\equiv 0\equiv Q_{\mu\nu\lambda}$), the rest of the terms on the right-hand side of (\ref{nmtRay}) vanish identically. Then, setting $n=4$, we recover the standard form of the Raychaudhui equation (e.g.~see~\cite{TCM} and also keep in mind that $a_{\mu}\equiv A_{\mu}$, with $a_{\mu}u^{\mu}=0= A_{\mu}u^{\mu}$, and that $\xi_{\mu}\equiv0$ when metricity holds).

The Raychaudhuri equation derived in this sections, as well as its reduced expressions given in the following sections (see \S~\ref{ssCPT} and \S~\ref{ssCPN-M} next), is a purely geometrical relation. As yet, no matter sources have been introduced and no assumption has been made about the nature of the gravitational field. One could add physical context to these geometrical expressions by introducing a set of field equations, like the Einstein, or the Einstein-Cartan, equations for example. In principle, Eq.~(\ref{nmtRay}) should be compatible with any geometrical theory of gravity.

Finally, it is worth stressing that the torsion terms in the second line on the right-hand side of Eq.~(\ref{nmtRay}) share a certain ``resemblance'' with the non-metricity terms seen in the fourth line of the same formula. This analogy, which is likely to reflect a deeper interconnection between torsion and non-metricity, will become more apparent in \S~\ref{ssVT} and \S~\ref{ssWN-M} below.

\subsection{The case of pure torsion}\label{ssCPT}
The terms in the second line on the right-hand side of Eq.~(\ref{nmtRay}) are purely torsional in nature, with the exception of the first which has a additional contribution from the non-metricity of the space (through the inner product $a_{\mu}u^{\mu}$, which vanishes when metricity holds). Then, when dealing with a $n$-dimensional spacetime that has nonzero torsion but satisfies the metricity condition, expression (\ref{nmtRay}) reduces to
\begin{eqnarray}
{\Theta}^{\prime}&=& -{1\over n-1}\,\Theta^2- R_{\mu\nu}u^{\mu}u^{\nu}- 2\left(\sigma^2-\omega^2\right)+ {\rm D}^{\mu}A_{\mu}+ A^{\mu}A_{\mu} \nonumber\\ &&+{2\over n-1}\,\Theta S_{\mu}u^{\mu}+ 2S_{\mu\nu\lambda}u^{\mu} \left(\sigma^{\nu\lambda}+\omega^{\nu\lambda}\right)+ 2S_{\mu\nu\lambda}A^{\mu}u^{\nu}u^{\lambda}\,,  \label{tRay}
\end{eqnarray}
with the prime indicating differentiation with respect to proper time (see \S~\ref{ssT-LOs} earlier). Applying the above to a 4-dimensional spacetime, one recovers the Raychaudhuri equation of the Riemann-Cartan geometry derived in~\cite{PTB}. Note that, when doing the aforementioned identification, one should also take into account the differences in the definitions of the torsion tensor and of the torsion vector between the two studies.

Following (\ref{tRay}), torsion affects the convergence/divergence of a timelike congruence in a variety of ways, which depend on whether these worldlines are geodesics or not, as well as on whether they have nonzero shear or vorticity. The most straightforward effect of torsion propagates via the first term in the second line on the right-hand side of the above. More specifically, torsion enhances/inhibits the expansion/contraction of the worldline congruence depending on the sign of the inner product ($S_{\mu}u^{\mu}$) between the torsion vector and the $n$-velocity (i.e.~on the relative orientation of the two vector fields -- see also~\cite{PTB} for further discussion).

As we mentioned in the previous section, Eq.~(\ref{tRay}) is of purely geometrical nature, since no matter fields have been introduced yet. In order to investigate the effects of gravity, we need to relate both the Ricci tensor and the torsion tensor to the material component of the spacetime. This can be done by means of, say, the Einstein-Cartan and the Cartan field equations~\cite{PTB}.

\subsection{The case of pure non-metricity}\label{ssCPN-M}
Finally, the terms seen in lines three to six on the right-hand side of (\ref{nmtRay}) are due to the non-metricity of the space, with the last of them carrying a torsional contribution as well. Therefore, in the presence of non-metricity but in the absence of torsion, we may write
\begin{eqnarray}
\dot{\Theta}&=& -{1\over n-1}\,\Theta^2- R_{\mu\nu}u^{\mu}u^{\nu}- 2\left(\sigma^2-\omega^2\right)+ {\rm D}^{\mu}a_{\mu}+ {1\over\ell^2}\,a_{\mu}A^{\mu} \nonumber\\ &&-{1\over\ell^2}(a_{\mu}u^{\mu})^{\cdot}- {2\Theta\over\ell^2(n-1)}\,a_{\mu}u^{\mu}+ {n-2\over\ell^4(n-1)}(a_{\mu}u^{\mu})^2+ {2\over\ell^2}\,a_{\mu}\xi^{\mu}- \dot{\tilde{Q}}_{\mu}u^{\mu} \nonumber\\ &&+{1\over n-1} \left(\Theta+{1\over\ell^2}\,a_{\nu}u^{\nu}\right) (Q_{\mu}-\tilde{Q}_{\mu})u^{\mu}-Q_{\mu\nu\lambda} (\sigma^{\mu\nu}+\omega^{\mu\nu}) u^{\lambda}- {1\over\ell^2}\,Q_{\mu\nu\lambda}u^{\mu}u^{\nu} (a^{\lambda}+\xi^{\lambda}) \nonumber\\ &&+Q_{\mu\nu\lambda}u^{\mu}\sigma^{\nu\lambda}+ {1\over\ell^2}\,Q_{\mu\nu\lambda}(u^{\mu}\xi^{\nu}+a^{\mu}u^{\nu}) u^{\lambda}+ u^{\mu}u^{\nu}\nabla^{\lambda}Q_{\mu\nu\lambda}+ Q_{\mu}{}^{\lambda\beta}Q_{\beta\lambda\nu}u^{\mu}u^{\nu}\,.  \label{nmRay}
\end{eqnarray}
Here, in contrast to Eq.~(\ref{tRay}), the overdot implies differentiation in terms of the affine parameter (i.e.~relative to $\lambda$ -- see \S~\ref{ssT-LOs}). According to the above, the implications of non-metricity for the convergence/divergence of a timelike congruence are multiple and not straightforward to decode. Similarly to the case of pure torsion seen before, the most transparent effects are those depending on the orientation of the non-metricity vectors and their derivatives (i.e.~$Q_{\mu}$, $\tilde{Q}_{\mu}$ and $\dot{\tilde{Q}}_{\mu}$) relative to the $u_{\mu}$-field.

Before closing this section, we should point out that the Raychaudhuri formulae given in expressions (\ref{nmtRay})-(\ref{nmRay}), are purely geometrical relations, which acquire physical relevance after the energy-momentum and the hyper-momentum tensors are introduced. The former gives rise to spacetime curvature, while the latter leads to both torsion and non-metricity through the field equations and the Palatini equations respectively. Also note that the nature of the observers' worldlines, namely of the curves tangent to the $n$-velocity vector $u_{\mu}$, has so far been left unspecified. Assuming, for example, motion along autoparallel curves the path-acceleration vanishes (i.e.~$A_{\mu}=0$ -- see \S~\ref{ssPHn-A} earlier).

\section{Characteristic cases}\label{sCCs}
According to Eq.~(\ref{nmtRay}), torsion and non-metricity affect the mean expansion/contraction of the host spacetime in a variety of intricate ways. In this section we will try to reveal the role of torsion and non-metricity in some characteristic cases.

\subsection{Vectorial torsion}\label{ssVT}
The kinematic effects of torsion (and spin) have been investigated primarily within the framework of the Einstein-Cartan theory. Assuming that the metricity condition holds (i.e.~setting $Q_{\mu\nu\lambda}=0)$, let us consider the case of vectorial torsion with
\begin{equation}
S_{\mu\nu\lambda}= \frac{2}{n-1}\,S_{[\mu}g_{\nu]\lambda}\,,  \label{vtrs}
\end{equation}
where $S_{\mu}=S_{\mu\nu}{}^{\nu}$ defines the associated torsion vector (e.g.~see~\cite{O}). Note that in this case the connection is $\Gamma^{\lambda}{}_{\mu\nu}=\tilde{\Gamma}^{\lambda}{}_{\mu\nu}+ 2(S_{\mu}\delta^{\lambda}{}_{\nu}-S^{\lambda}g_{\mu\nu})/(n-1)$, with $\tilde{\Gamma}^{\lambda}{}_{\mu\nu}$ being the Christoffel symbols. Then, the second-last term on the right-hand side of (\ref{tRay}) vanishes, while the last one reduces to $2S_{\mu\nu\lambda}A^{\mu}u^{\nu}u^{\lambda}=-2S_{\mu}A^{\mu}/(n-1)$. As a result, the Raychaudhuri equation recasts into
\begin{eqnarray}
\Theta^{\prime}&=& -{1\over n-1}\,\Theta^2- R_{\mu\nu}u^{\mu}u^{\nu}- 2\left(\sigma^2-\omega^2\right)+ {\rm D}_{\mu}A^{\mu}+ A_{\mu}A^{\mu} \nonumber\\ &&+{2\over n-1}\,\Theta S_{\mu}u^{\mu}- {2\over n-1}\,S_{\mu}A^{\mu}\,.  \label{vtRay1}
\end{eqnarray}
Consequently, the effects of vectorial torsion on the mean expansion/contraction of the host spacetime, depend on the orientation of the torsion vector relative to the observer's velocity and acceleration. In particular, when $S_{\mu}$ is purely timelike, we have $S_{\mu}A^{\mu}=0$ (recall that $A_{\mu}u^{\mu}=0$ when metricity holds). For purely spacelike torsion vector, on the other hand, $S_{\mu}u^{\mu}=0$.

Suppose now that the $u_{\mu}$-field is tangent to a congruence of autoparallel curves in a 4-dimensional spacetime (i.e.~set $A_{\mu}=0$ and $n=4$). Assume also a Ricci-flat (i.e.~empty) spacetime with homogeneous and isotropic spatial hypersurfaces (i.e.~set $R_{\mu\nu}=0=\sigma_{\mu\nu}=\omega_{\mu\nu}$). In such an FRW-like environment, expression (\ref{vtRay1}) reduces to
\begin{equation}
\Theta^{\prime}= -{1\over3}\,\Theta^2+ {2\over3}\,\Theta S_{\mu}u^{\mu}= -{1\over3}\,\Theta \left(\Theta-2S_{\mu}u^{\mu}\right)\,,  \label{vtRay2}
\end{equation}
while the torsion vector becomes purely timelike (to preserve the isotropy of the 3-space).\footnote{In FRW-type models with torsion the associated torsion tensor is conveniently given by the ansatz $S_{\mu\nu\lambda}=2\phi u_{[\mu}h_{\nu]\lambda}$, where $\phi$ is a scalar function that depends only on time~\cite{Ts}. It is then straightforward to show that $S_{\mu}= S_{\mu\nu}{}^{\nu}=3\phi u_{\mu}$. Note that the aforementioned torsion ansatz is a special case of definition (\ref{vtrs}).} Therefore, the vectorial torsion increases or decreases the rate of the mean expansion/contration of a timelike congruence, depending on whether the torsion vector is (respectively) parallel or antiparallel to the $u_{\mu}$-field. We may take a qualitative look by employing the relation $\Theta= \tilde{\Theta}+2S_{\mu}u^{\mu}$, where $\tilde{\Theta}$ represents the purely Riemannian (i.e.~the torsionless) counterpart of the expansion/contraction scalar (e.g.~see~\cite{PTB}). Recalling that $\tilde{\Theta}/3=a^{\prime}/a$, with $a=a(\tau)$ being the associated scale factor, solving the above relation for $S_{\mu}u^{\mu}$ and then substituting the resulting expression into the right-hand side of Eq.~(\ref{vtRay2}), we find that $\Theta^{\prime}/\Theta= -a^{\prime}/a$. The latter integrates immediately to give
\begin{equation}
\Theta= \Theta_0\left({a_0\over a}\right)\,,  \label{vtTheta}
\end{equation}
with the zero suffix marking a given initial time. According to the above solution, in an expanding spacetime (with $\Theta_0>0$), we find that $\Theta\rightarrow 0^{+}$ at late times (i.e.~as $a\rightarrow+\infty$). When dealing with contracting models, on the other hand, we have $\Theta_0<0$. In this case, solution (\ref{vtTheta}) ensures that $\Theta\rightarrow-\infty$ as $a\rightarrow0^{+}$. In the former example the expansion comes (asymptotically) to a halt, while in the latter the (autoparallel) worldline congruence focuses at a point.\footnote{Generally speaking, a singularity in the volume scalar (i.e.~$\Theta\rightarrow-\infty$) means that caustics will develop in the worldline congruence and does not necessarily imply a singularity in the spacetime structure~\cite{Wa}.}

Not surprisingly, the quantitative effect of vectorial torsion on the mean kinematics of the host spacetime depends on the specific form of the associated torsion vector. We can demonstrate this dependence by solving Eq.~(\ref{vtTheta}) for the cosmological scale factor ($a=a(\tau)$). More specifically, using the result $a\Theta= a_0\Theta_0=$~constant and the splitting $\Theta= \tilde{\Theta}+2S_{\mu}u^{\mu}$, of the volume scalar into its purely Riemannian and torsional parts, we arrive at
\begin{equation}
a^{\prime}+ {2\over3}\left(S_{\mu}u^{\mu}\right)\,a= a_0\Theta_0= \mathcal{C}_0\,.  \label{adot}
\end{equation}
Keeping in mind that the torsion vector is purely timelike due to the spatial symmetry and homogeneity of the Friedmann-like spacetimes, the above accepts the solution
\begin{equation}
a= a(\tau)= {\rm e}^{-{2\over3}\int S_{\mu}u^{\mu}{\rm d}\tau} \left[\mathcal{C}_1+\mathcal{C}_0\int{\rm e}^{{2\over3}\int S_{\mu}u^{\mu}{\rm d}\tau}{\rm d}\tau\right]\,,  \label{vta}
\end{equation}
where the integration constant $\mathcal{C}_0$ and $\mathcal{C}_1$ are decided by the initial conditions. Therefore, in metric-compatible FRW-type spacetimes with nonzero torsion, the scale factor evolution is decided by the product $S_{\mu}u^{\mu}$, namely by the orientation of the torsion vector relative to the $u_{\mu}$-field. Interestingly, solution (\ref{vta}) also allows for the exponential increase of the scale factor. This can happen, for example, when the scalar $S_{\mu}u^{\mu}$ equals a negative constant.

\subsection{Weyl non-metricity}\label{ssWN-M}
The Weyl non-metricity is also of vectorial form, since $Q_{\mu\nu\lambda}=Q_{\mu}g_{\nu\lambda}/n$, with $Q_{\mu}=Q_{\mu\nu}{}^{\nu}$ representing the associated Weyl vector (see definition (\ref{nmvecs}a) in \S~\ref{ssSTTN-M} earlier). Then, $\tilde{Q}_{\mu}=Q^{\nu}{}_{\nu\mu}=Q_{\mu}/n$ (see definition (\ref{nmvecs}b)), leaving only one independent non-metricity vector. Here, the connection is $\Gamma^{\lambda}{}_{\mu\nu}= \tilde{\Gamma}^{\lambda}{}_{\mu\nu}+ (2\delta^{\lambda}{}_{(\mu}Q_{\nu)}-Q^{\lambda}g_{\mu\nu})/2n$, where $\tilde{\Gamma}^{\lambda}{}_{\mu\nu}$ are the Christoffel symbols. Therefore, assuming zero torsion, Weyl non-metricity and confining to autoparallel curves (i.e.~those with $A_a=0$), the Raychaudhuri equation (see expression (\ref{nmRay}) in \S~\ref{ssCPN-M}) reduces to
\begin{eqnarray}
\left(\Theta-2\,{\dot{\ell}\over\ell}\right)^{\cdot}= -{1\over n-1} \left(\Theta-2\,{\dot{\ell}\over\ell}\right)^2- R_{\mu\nu}u^{\mu}u^{\nu}- 2\left(\sigma^2-\omega^2\right)\,.  \label{WnmRay1}
\end{eqnarray}
Note that, in deriving the above, we have utilised the relation $a_{\mu}=-(Q_{\nu}u^{\nu}/n)u_{\mu}$, which connects the hyper acceleration to the Weyl vector in the case of autoparallel motion (see Eq.~(\ref{Aa}) in \S~\ref{ssPHn-A}). Then, one can immediately obtain the auxiliary results $a_{\mu}u^{\mu}=-2\ell\dot{\ell}$ and $Q_{\mu}u^{\mu}= -2n\dot{\ell}/\ell$, which also hold for Weyl non-metricity and for zero path acceleration. In addition, we have $\sigma_{\mu\nu}g^{\mu\nu}= \sigma_{\mu}{}^{\mu}=0$ and $\xi_{\mu}u^{\mu}=a_{\mu}u^{\mu}$ by construction.

Confining to a 4-dimensional spacetime and assuming an autoparallel congruence that is also irrotational and shear-free, namely setting $n=4$ and $\omega=0=\sigma$ in Eq.~(\ref{WnmRay1}), the latter leads to
\begin{eqnarray}
\left(\Theta-2\,{\dot{\ell}\over\ell}\right)^{\cdot}+ {1\over3} \left(\Theta-2\,{\dot{\ell}\over\ell}\right)^2\leq 0\,,  \label{WnmRay2}
\end{eqnarray}
provided that $R_{\mu\nu}u^{\mu}u^{\nu}\geq0$. This last constraint on the Ricci tensor may be seen as the generalisation of the familiar ``weak energy condition'' to spacetimes with (Weyl) non-metricity. It is then straightforward to show (e.g.~see~\cite{Wa} for details) that (\ref{WnmRay2}) integrates to
\begin{equation}
\left(\Theta-2\,{\dot{\ell}\over\ell}\right)^{-1}\geq \left[\Theta_0-2\left({\dot{\ell}\over\ell}\right)_0\right]^{-1}+ {1\over3}\,\lambda\,,  \label{intWnmRay}
\end{equation}
with the zero suffix marking a given initial affine value. Starting from the above and following~\cite{Wa}, we deduce that $\Theta-2\dot{\ell}/\ell\rightarrow-\infty$ within finite affine length (i.e.~for $\lambda\leq [\Theta_0-2(\dot{\ell}/\ell)_0]/3$), assuming that $\Theta_0-2(\dot{\ell}/\ell)_0<0$ initially. Put another way, provided that $\Theta_0<2(\dot{\ell}/\ell)_0$, the volume scalar of the congruence will develop a caustic singularity (i.e. $\Theta\rightarrow-\infty$), unless $\dot{\ell}/\ell\rightarrow+\infty$ simultaneously. An interesting deviation from the standard Riemannian studies is that, when $(\dot{\ell}/\ell)_0>0$, caustic formation seems now possible even for initially expanding congruences, namely for those with $0<\Theta_0<2(\dot{\ell}/\ell)_0$.

Before attempting to solve Eq.~(\ref{WnmRay1}), it helps to decompose the volume scalar into a purely Riemannian component and the non-metricity contribution. Recalling that $\Theta=g^{\mu\nu}\nabla_{\nu}u_{\mu}$ by definition, we find that that $\Theta=\nabla_{\mu}u^{\mu}- Q_{\mu}u^{\mu}/n=\nabla_{\mu}u^{\mu}+2\dot{\ell}/\ell$ in the case of Weyl non-metricity. In addition, we have $\nabla_{\mu}u^{\mu}=\tilde{\nabla}_{\mu}u^{\mu}+ Q_{\mu}u^{\mu}/2=\partial_{\mu}u^{\mu}+ \tilde{\Gamma}^{\mu}{}_{\nu\mu}u^{\nu}-\dot{\ell}/\ell$, with the latter equality also holding for Weyl non-metricity. Combining all the above gives
\begin{equation}
\Theta= \partial_0\ell+ \tilde{\Gamma}^{\mu}{}_{0\mu}\ell+ {\dot{\ell}\over\ell}= \tilde{\Gamma}^{\mu}{}_{0\mu}\ell+ 2\,{\dot{\ell}\over\ell}\,,  \label{WnmTheta1}
\end{equation}
since $u^{\mu}=\delta^{\mu}{}_0\ell$ and $\partial_0\ell= \ell^{\prime}=\dot{\ell}/\ell$. Finally, keeping in mind that $\tilde{\Gamma}^0{}_{00}=0$ and $\tilde{\Gamma}^1{}_{01}= \tilde{\Gamma}^2{}_{02}=\tilde{\Gamma}^3{}_{03}=a^{\prime}/a$ in a flat FRW spacetime, we arrive at
\begin{equation}
\Theta= 3\,{\dot{a}\over a}- {\dot{\ell}\over\ell}\,.  \label{WnmTheta2}
\end{equation}

Let us now apply Eq.~(\ref{WnmRay1}) to a congruence of irrotational and shear-free autoparallel curves ``living'' in a Ricci-flat 4-dimensional spacetime. Then, a straightforward integration of the remaining differential equation leads to
\begin{equation}
\Theta- 2\,{\dot{\ell}\over\ell}= \left({1\over3}\,\lambda+\mathcal{C}\right)^{-1}\,, \label{WnmTheta3}
\end{equation}
where the integration constant ($\mathcal{C}$) depends on the initial conditions. In addition, keeping in mind that $\Theta=3\dot{a}/a-\dot{\ell}/\ell$ (see Eq,~(\ref{WnmTheta2}) above), the left-hand side of (\ref{WnmTheta3}) reads $[\ln(a/\ell)^3]^{\cdot}$ and we arrive at the following expression
\begin{equation}
a= a(\lambda)= \ell\left(\mathcal{C}_1+\mathcal{C}_2\lambda\right)\,,
\label{Wnma1}
\end{equation}
for the scale factor in terms pf the affine parameter. To proceed further, recall that $\ell^2=\int Q_{\mu\nu\lambda}u^{\mu}u^{\nu}u^{\lambda}{\rm d}\lambda+ \mathcal{C}$ when dealing with autoparallel curves (see \S~\ref{ssPHn-A} earlier). Therefore, for Weyl non-metricity in a 4-dimensional spacetime, we find
\begin{equation}
\ell= \ell_0{\rm e}^{-{1\over8}\int Q_{\mu}u^{\mu}{\rm d}\lambda}\,.  \label{Wnmell}
\end{equation}
Furthermore, substituting the above expression into Eq.~(\ref{tau-lambda}) and integrating leads to
\begin{equation}
\lambda= \pm{1\over\ell_0}\int{\rm e}^{-{1\over8}\int Q_{\mu}u^{\mu}{\rm d}\lambda}\,.  \label{Wnmlambda}
\end{equation}
Finally, on using the auxiliary relations (\ref{Wnmell}) and (\ref{Wnmlambda}), expression (\ref{Wnma1}) recasts into
\begin{equation}
a= a(\lambda)= {\rm e}^{-{1\over8}\int Q_{\mu}u^{\mu}{\rm d}\lambda} \left(\mathcal{C}_3+\mathcal{C}_4\int{\rm e}^{{1\over8}\int Q_{\mu}u^{\mu}{\rm d}\lambda}\right)\,.
\label{Wnma2}
\end{equation}
The above solution provides the scale factor in terms of the affine parameter of an autoparallel congruence of irrotational and shear-free worldlines, which reside in a 4-dimensional, Ricci-flat spacetime ``equipped'' with Weyl non-metricity. In close analogy with the case of pure (vectorial) torsion (see \S~\ref{ssVT} before), when the host spacetime is FRW-like, the non-metricity vector is purely timelike. Also, the scale-factor evolution is decided by the scalar product $Q_{\mu}u^{\mu}$, that is by the orientation (parallel or antiparallel) of the non-metricity vector relative to the $u_{\mu}$-field. What is most intriguing, however, is that expression (\ref{Wnma2}) is formally identical to the pure-torsion solution (\ref{vta}). In fact, the two expressions are indistinguishable, provided we make the simple exchange $Q_{\mu}\leftrightarrow16S_{\mu}/3$ and interchange proper time with the affine parameter in the related integrals.\footnote{In analogy with its pure-torsion analogue derived in the previous section, solution (\ref{Wnma2}) also allows for the exponential increase of the scale factor. Similarly to the torsion case seen in \S~\ref{ssVT}, this could happen when the scalar product $Q_{\mu}u^{\mu}$ equals a negative constant. This result, which requires further scrutiny, seems to support earlier claims made in the literature about the theoretical possibility of a non-metricity driven inflation~\cite{St}.} This apparent ``duality'' between torsion and non-metricity has been observed and reported in earlier works as well~\cite{BSS}. Here, we see that in highly symmetric (Friedmann-like) spacetimes where only the vector components of torsion and non-metricity survive, the effects of the aforementioned two geometrical agents are phenomenologically indistinguishable.

\section{Discussion}\label{sD}
Classical general relativity combines theoretical elegance and observational success at the highest level. Nevertheless, modifications/extensions of Einstein's theory have been proposed and investigated ever since relativity was introduced in the early years of the last century. The motivation behind these efforts are multiple, ranging from the quest for quantum gravity and the existence of singular solutions for key relativistic equations, to the awareness of the intrinsic limitations of the theory and its apparent inability to explain certain observations. Violating the metricity condition and including spacetime torsion have long been suggested as possible ways of ``improving'' standard general relativity. Technically speaking the non-compatibility of the metric and the asymmetry of the connection imply that the latter is no longer uniquely defined by the former. In other words, the metric and the connection are treated as independent geometrical fields, an approach that is often referred to as the ``Palatini formalism'', although a more precise terminology is metric-affine formalism.

Historically speaking, non-metricity was first introduced to unify gravity with electromagnetism and torsion to incorporate the nonzero spin of the matter into the gravitational field. In the literature there are several suggestions, as well as a debate, on the possibility of experimentally testing torsion~\cite{MTGC}. Although less frequent, there is also discussion on potentially measurable effects from non-metricity~\cite{FKX}. In this work we have considered a generalised spacetime with n-dimensions, nonzero torsion and general non-metricity. Our aim was to study the mean kinematics of timelike worldlines and see how these are affected by the aforementioned two extra spacetime features. We did so, by employing and extending the 1+3 covariant formalism, which combines both mathematical compactness and physical clarity, to spaces with torsion and non-metricity. After adapting the covariant approach to the new environment and clarifying several subtle issues, we derived and provided the most general (to the best of our knowledge) version of Raychaudhuri's formula. The latter is known to monitor the mean kinematics of timelike observers and has been the key formula for studying self-gravitating media.

Not surprisingly, the introduction of extra degrees of freedom into the host spacetime added several new effects to the Raychaudhuri equation. This in turn made the kinematics of the residing observers considerably more involved and therefore more difficult to decode. Nevertheless, by treating torsion and non-metricity separately and by confining to highly symmetric (Friedmann-like) spacetimes, we were able to obtain both qualitative results and analytical solutions. In particular, assuming vectorial torsion and Weyl non-metricity, we found that the solutions of the associated Raychaudhuri equations were formally identical. More specifically, it was shown that one could recover the former solution from the latter (and vice versa), by merely imposing a simple ansatz between the torsion and the Weyl vectors. Analogous reports of such a ``duality'' relation between these two geometrical agents are not uncommon in the literature. We attribute ours to the high symmetry of the host spacetimes, which appears to make the effects of torsion and non-metricity macroscopically indistinguishable.\newline

\textbf{Acknowledgements:} We would like to thank Lavinia Heisenberg, Alan Kostelecky, Iarley Lobo and Simone Speziale for their helpful comments. CGT acknowledges support from Clare Hall College and by DAMTP at Cambridge University, where part of this work was conducted.

\end{document}